\documentclass[showpacs,twocolumn,aps]{revtex4}
\usepackage{amssymb}
\usepackage{amsmath}
\usepackage{graphicx}
\usepackage{lscape}
\usepackage{booktabs}
\begin{document}
\title{Parton rescattering effect on particle production in ultra-relativistic
       p+p collisions}
\author{Yu-Liang Yan$^{1}$, Bao-Guo Dong$^{1,2}$, Dai-Mei Zhou$^{3}$, Xiao-Mei
Li$^{1}$, Hai-Liang Ma$^{1}$, Ben-Hao Sa$^{1,3,4}$\footnote{Corresponding
author: sabh@ciae.ac.cn}}
\address{ 1 China Institute of Atomic Energy, P.O. Box 275 (18), Beijing
102413, China \\
2 Center of Theoretical Nuclear Physics, National Laboratory of
Heavy Ion Collisions, Lanzhou 730000, China\\
3 Institute of Particle Physics, Huazhong Normal University, Wuhan
430079, China\\
4 CCAST (World Laboratory), P. O. Box 8730 Beijing 100080, China}

\begin{abstract}
The parton rescattering effect on the charged particle production
in ultra-relativistic p+p collisions is studied by the parton and
hadron cascade model, PACIAE, based on PYTHIA. We have calculated
charged particle pseudorapidity density ($dN_{ch}/d\eta$) at
mid-rapidity and the pseudorapidity distribution in inelastic and
non-single diffractive p+p collisions at $\sqrt s$=200, 900, 5500,
and 14000 GeV with the PYTHIA and PACIAE models. The calculated results 
of $\sqrt s$=900 GeV are well compared with the ALICE data. The calculated 
$dN_{ch}/d\eta$ as a function of center-of-mass energy well meets with the
experimental data as well. Comparing the PYTHIA results (without parton
rescattering) with the PACIAE results (with parton rescattering),
it turned out that the parton rescattering effect plays an
important role and this effect increases with increasing center-of-mass
energy.

\end{abstract}

\maketitle

\section {INTRODUCTION}
The jets, as remnants of hard-scattered quarks and gluons, have
been investigated to explore the properties of partons
\cite{star1}. Studying the hot QCD matter via jet quenching
(partons lose energy on their way through the QCD medium) has now
become an important role of jets and has been realized at RHIC
\cite{star2,salu}. Experimentally extracting the partonic
observables from final state hadrons with reconstruction method is
now highly interesting \cite{star3,poch}.

Another way to study the properties of partons is introduced
recently. The $p_T$ distribution of effective $u$ ($d$) and $s$
quarks are extracted from the ratio of $\Xi(p_T/3)/\phi(p_T/2)$
and $\Omega(p_T/3)/\phi(p_T/2)$ in the Au+Au collisions at
$\sqrt{s_{{NN}}}$=200 GeV , respectively \cite{huan,chen}. In this
extraction it is assumed that the hadron's $p_T$ is composed of
its effective constituent quark's $p_T/n$ (n is the number of
effective constituent quarks in the hadron) and the different
quark and anti-quark have the same $p_T$ distribution.

The measurement of elliptic flow parameter $v_2$ relies on the
analysis techniques which require high event multiplicity.
Therefore, no report about the $v_2$ measurement in p+p collisions
is published by now because of low multiplicity. However, in
recent papers \cite{boze,casa,chau} it is argued that the elliptic
flow parameter in high multiplicity p+p collisions at the LHC
energy ($\sqrt s$=14000 GeV) may be measurable. Similarly, one could 
expect the formation probability of quark-gluon mater (QGM) in the p+p
collisions at RHIC energy ($\sqrt s$=200 GeV) is very low \cite{arme}. The 
parton rescattering effect on the final state hadrons in p+p collisions at 
RHIC energy is weak as well \cite{yan}. However, those probability and 
effect may be visible in the p+p collisions at LHC or higher energy 
\cite{yan}.

The study of ultra-relativistic p+p collisions is a door sill
toward the nucleus-nucleus collisions at the same energy. Because
the nucleus-nucleus collision can be decomposed into
nucleon-nucleon collisions and the nucleon-nucleon collision can
be well described by the perturbative QCD (pQCD) due to its
relatively clear and simple physics. The results of p+p collisions
are a worthy reference for nucleus-nucleus collisions. Thus the
ALICE first successful measurements for p+p collisions at $\sqrt
s$=900 GeV \cite{alic1} not only demonstrate ``the LHC and its
experiments have finally entered the phase of physics
exploitation" but also light the way studying Pb+Pb collisions at
LHC energies. It also means the copious measurements in p+p
collisions at LHC energies are forthcoming and will supply
opportunities for the investigations of the issues mentioned
above.

As a dedication to the ALICE first measurements, in this paper the
parton and hadron cascade model PACIAE, based on PYTHIA, is used
to analyze the ALICE first measurements, especially to explore the
parton rescattering effect on the final state hadron in the
ultra-relativistic p+p collisions. This investigation is not only
a quick response to the ALICE first measurements but also a well
preparation for the further studies in p+p and Pb+Pb collisions at
LHC energies.

\section {MODELS}
The parton and hadron cascade model, PACIAE \cite{sa}, is based on
PYTHIA \cite{soj2}. PYTHIA is a model for high energy
hadron-hadron (hh) collisions. In the PYTHIA model a hh collision
is decomposed into the parton-parton collisions. A hard
parton-parton collision is described by the lowest leading order
perturbative QCD (LO-pQCD) parton-parton interactions modified by
the parton distribution function in a hadron. The soft
parton-parton collision is considered empirically. Because the
initial- and final-state QCD radiations are considered in
parton-parton interactions, the consequence of a hh collision is a
partonic multijet configuration composed of di-quarks
(anti-diquarks), quarks (anti-quarks), and gluons. This is
followed by the string construction and the string fragmentation.
So one obtains a final hadronic state for a hh (p+p) collision.

The differences between the PACIAE model and PYTHIA model, for p+p
collision, are as follows:
\begin{itemize}
\item The string fragmentation mentioned above is switch-off
temporarily in the PACIAE model. Therefore, if the di-quarks
(anti-diquarks) are broken randomly into quarks (anti-quarks) the
consequence of a hh (p+p) collision is a configuration of quarks,
anti-quarks, and gluons. As such the partonic initialization stage
for a p+p collision is realized. \item Then the rescattering among
partons is introduced and performed by the Monte Carlo method
until partonic freeze-out (no more parton-parton interaction at
all). In this parton rescattering stage the LO-pQCD parton-parton
interaction cross sections (2 $\rightarrow$ 2) are employed
\cite{comb}. \item In the next hadronization stage, the partonic
matter formed after parton rescattering is hadronized by the Lund
string fragmentation regime \cite{soj2} and/or coalescence model
\cite{sa}. \item The consequent hadronic matter suffers hadronic
rescattering. We deal with hadronic rescattering as usual two-body
collision \cite{sa1}, until hadronic freeze-out (hh collision pair
is exhausted). This is the hadronic rescattering stage.
\end{itemize}
\section {PARTON RESCATTERING EFFECT ON PARTICLE PRODUCTION}
As we aim at the physics behind the experimental data rather than
reproducing the data, in all calculations the model parameters are
fixed at their default values, except the $k$ factor, considering
the higher order and un-perturbative QCD corrections, is assumed
to be equal to 3. Both inelastic (INEL) p+p and non-single
diffractive (NSD) p+p collisions are calculated with and without
parton rescattering at $\sqrt s$=200, 900, 5500, and 14000 GeV.
The calculated results without parton rescattering will be
indicated as PYTHIA and the results with parton rescattering as
PACIAE. In addition, a similar calculations are also performed for
the p+\=p collisions at $\sqrt s$= 900 GeV.

\begin{widetext}
\begin{center}
\begin{table}[htbp]
\caption{Charged particle pseudorapidity density at mid-rapidity
($|\eta|<$ 0.5) calculated by the PYTHIA and PACIAE models for p+p
collisions at various center-of-mass energies. The ALICE p+p and
UA5 p+\=p data \cite{alic1,ua5} at $\sqrt s$=900 GeV are given as
well. In the ALICE data the first error is statistical error and
the second is systematic one.}
%\footnotesize{
%\scriptsize{
\begin{tabular}{cccccccc}
\hline\hline
         & & p+p  & & & &p+\=p &  \\
\cmidrule[0.25pt](l{0.05cm}r{0.05cm}){2-5}
\cmidrule[0.25pt](l{0.05cm}r{0.05cm}){6-8}
  & ALICE& 320 tune$^1$& PYTHIA& PACIAE& UA5& PYTHIA& PACIAE \\
  INEL& 3.10 $\pm$ 0.13 $\pm$ 0.22& 2.46& 2.59 & 2.80& 3.09 $\pm$ 0.05 & 2.70
  &2.94    \\
  NSD & 3.51 $\pm$ 0.15 $\pm$ 0.25& 3.02& 3.15 & 3.33& 3.43 $\pm$ 0.05 & 3.30
  &3.50    \\
\hline\hline \multicolumn{8}{l}{$^1$ The PYTHIA results of the
Perugia (320) tune taken from \cite{skan}.}
\end{tabular}
\label{mul}
\end{table}
\end{center}
\end{widetext}

Table I gives the calculated charged particle pseudorapidity
density at mid-rapidity ($|\eta|<$0.5), $dN_{ch}/d\eta$, together
with the ALICE data \cite{alic1} and UA5 data \cite{ua5}. One sees
in this table that:
\begin{itemize}
\item The PYTHIA results of $dN_{ch}/d\eta$ in this paper are
within the range of three PYTHIA tunes \cite{alic1}: from 2.33
\cite{albr} to 2.99 \cite{mora} for INEL and from 2.83 \cite{albr}
to 3.68 \cite{mora} for NSD. Our PYTHIA results are close to the
Perugia (320) tune \cite{skan}. 
\item Both the ALICE and UA5 data are better reproduced by PACIAE than 
PYTHIA. This means the parton rescattering effect is important in 
reproducing the experimental data. 
\item The difference in charged particle pseudorapidity
density between NSD p+\=p and p+p collisions at $\sqrt s$=900 GeV
is about 4.5\% in the PYTHIA calculations and 4.9\% in PACIAE.
They are larger than the range of 0.1-0.2\% resulted from
extrapolating the CERN ISR measurement \cite{isr} to $\sqrt s$
=900 GeV.
\end{itemize}

\begin{figure}[htbp]
\includegraphics[width=9cm,angle=0]{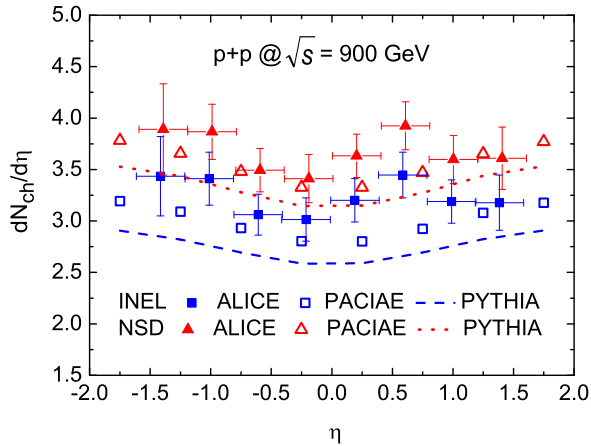}
\caption{(Color online) Charged particle pseudorapidity
distributions in INEL and NSD p+p collisions at $\sqrt s$=900 GeV.
The full squares and triangles are ALICE INEL and NSD data,
respectively, taken from \cite{alic1}. The open squares and
triangles are the PACIAE results. The dashed and dotted curves are
the PYTHIA results.} \label{pareff_pp1}
\end{figure}
\begin{figure}[htbp]
\includegraphics[width=9cm,angle=0]{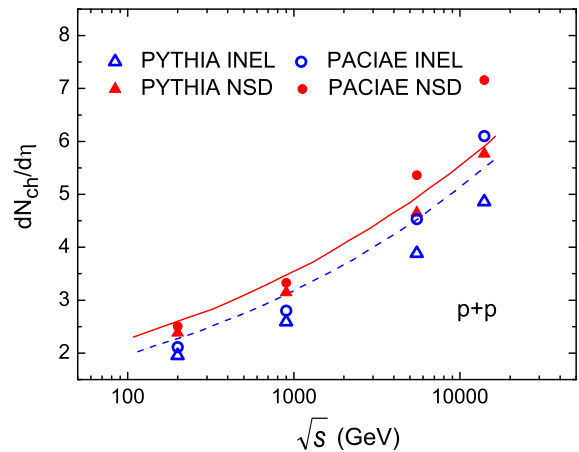}
\caption{(Color online) Charged particle pseudorapidity density at
mid-rapidity ($|\eta|<$0.5) in p+p collisions as a function of the
center-of-mass energy. The triangles and circles are the PYTHIA
and PACIAE results, respectively. The full and open symbols
indicate the NSD and INEL p+p collisions. The dashed and solid
lines indicate the fits to experimental data ($\sqrt{s}<$ 1900
GeV) by a power-law dependence on the energy for INEL and NSD
interactions, respectively \cite{alic1}.} \label{pareff_pp2}
\end{figure}

Figure \ref{pareff_pp1} gives ALICE data of charged particle
pseudorapidity distributions in p+p collisions at $\sqrt s$ =900
GeV \cite{alic1} together with the results from PYTHIA and PACIAE
calculations. We see in this figure that the ALICE data are better
reproduced by PACIAE than PYTHIA. Both PYTHIA and PACIAE results
show the similar shallow valley at $\eta\sim$0 to the ALICE data,
and the better symmetry relative to $\eta$=0.
\begin{figure}[htbp]
\includegraphics[width=9cm,angle=0]{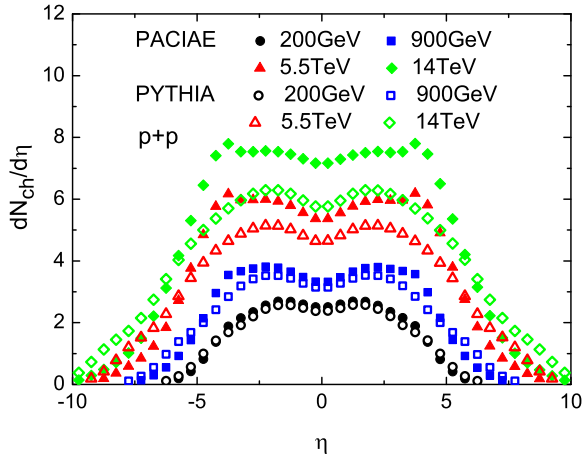}
\caption{(Color online) Charged particle pseudorapidity
distributions in NSD p+p collisions at $\sqrt s$=200 (circles),
900 (squares), 5500 (triangles), and 14000 GeV (diamonds)
calculated by PYTHIA (open symbols) and PACIAE (full symbols)
models, respectively.} \label{pareff_pp3}
\end{figure}

The calculated center-of-mass energy dependence of the charged
particle pseudorapidity density at mid-rapidity is compared with
the ALICE data in Fig. \ref{pareff_pp2}. One sees in this figure
that the power-law fit is better reproduced by PACIAE rather than
PYTHIA both for INEL and NSD. The PYTHIA results are below the
corresponding power-law fit lines for both INEL and NSD. The PACIAE NSD
results deviate from the corresponding power-law fit lines at
$\sqrt{s}\sim$2500 GeV. Then this deviation increases with
increasing center-of-mass energy. The role of parton rescattering
also increases with increasing center-of-mass energy. For
instance, the difference between PACIAE and PYTHIA NSD results (in
percentage) increases from 5.1\% at $\sqrt{s}$=200 GeV to 15.4\%
at 5500 GeV and then to 24.2\% at 14000 GeV.

We compare the charged particle pseudorapidity distributions
calculated by the PYTHIA and PACIAE models for NSD p+p collisions
at $\sqrt s$=200, 900, 5500, and 14000 GeV in Fig. \ref{pareff_pp3}. 
In this figure the open and full symbols are the PYTHIA and PACIAE 
results, respectively. The circles, squares, triangles, and diamonds 
indicate the results in $\sqrt s$=200, 900, 5500, and 14000 GeV p+p 
collisions, respectively. One sees the following features in this figure:
\begin{itemize}
\item The charged particle pseudorapidity density at mid-rapidity
and the width of center rapidity plateau increase with increasing
center-of-mass energy monotonously. 
\item In the center pseudorapidity region the PACIAE results are larger 
than PYTHIA. However, one sees the opposed situations in the outer
pseudorapidity region. This trend increases with increasing
center-of-mass energy. That means the parton rescattering effect
(the QGM formation probability) in p+p collisions increases with
increasing center-of-mass energy as well.
\end{itemize}

\section {CONCLUSIONS}
Using the PYTHIA model (without parton rescattering) and the parton
and hadron cascade model PACIAE (with parton rescattering) we have
investigated the charged particle productions in INEL and NSD p+p
collisions at $\sqrt s$=200, 900, 5500, and 14000 GeV as well as
p+\=p collisions at $\sqrt s$=900 GeV.

The calculated charged particle pseudorapidity density at
mid-rapidity ($|\eta|<$0.5) and the pseudorapidity distributions
in p+p and p+\=p collisions at $\sqrt s$=900 GeV well reproduce
the corresponding ALICE and UA5 data \cite{alic1,ua5}. The
calculated $dN_{ch}/d\eta$ at mid-rapidity as a function of
center-of-mass energy well meets with the power-law dependence
lines fitted to the experimental data in the $\sqrt s<$1900 GeV
region \cite{alic1}. Above this region the deviation of PACIAE NSD
results from NSD power-law dependence line increases with
increasing center-of-mass energy. The calculated charged particle
pseudorapidity distributions, similar to the ALICE data, show a
nearly plateau structure at mid-rapidity with a shallow valley at
center pseudorapidity. That plateau width increases with
increasing center-of-mass energy.

The parton rescattering effect plays an important role in
reproducing the experimental data. The parton rescattering effect
becomes larger and larger with increasing center-of-mass energy.
This also means the QGM formation probability in the p+p
collisions increases with increasing center-of-mass energy.
Therefore the QGM is possible to be formed in the early stage of
p+p collisions at ALICE or higher energy.

\acknowledgments

The financial support from NSFC (10635020, 10605040, 10705012,
10475032, 10975062, and 10875174) in China is acknowledged.

%\begin{references}

%\end{references}
\end{document}